\documentclass[aps,preprint]{revtex4}%
\usepackage{amsfonts}
\usepackage{amsmath}
\usepackage{amssymb}
\usepackage{graphicx}%
\begin{document}
\title{\textbf{Hawking radiation as tunneling derived from Black Hole Thermodynamics
through the quantum horizon}}
\author{Baocheng Zhang$^{a,b}$}
\email{zhangbc@wipm.ac.cn}
\author{Qing-yu Cai$^{a}$}
\author{Ming-sheng Zhan$^{a}$}
\affiliation{$^{a}$State Key Laboratory of Magnetic Resonances and Atomic and Molecular
Physics, Wuhan Institute of Physics and Mathematics, The Chinese Academy of
Sciences, Wuhan 430071, People's Republic of China}
\affiliation{$^{b}$Graduation University of Chinese Academy of Sciences, Beijing 100081,
Peope's Republic of China}
\author{}

\begin{abstract}
We show that the first law of the black hole thermodynamics can lead to the
tunneling probability through the quantum horizon by calculating the change of
entropy with the quantum gravity correction and the change of surface gravity
is presented clearly in the calculation. The method is also applicable to the
general situation which is independent on the form of black hole entropy and
this verifies the connection of black hole tunneling with thermodynamics
further. In the end we discuss the crucial role of the relation between the
radiation temperature and surface gravity in this derivation.\textbf{\ }

PACS classification codes: 04.70.Dy, 04.60.-m


\end{abstract}
\maketitle

\section{\textbf{Introduction}}

About 30 years ago, Hawking discovered \cite{swh75} that when considering
quantum effect black holes could radiate particles as if they were hot bodies
with the temperature $\kappa/2\pi$ where $\kappa$ was the surface gravity of
the black hole and explained \cite{hh76} the particles of radiation as
stemming from vacuum fluctuations tunneling through the horizon of the black
hole with Hartle together. But the semiclassical derivation of Hawking based
on the Bogoliubov transformation didn't have the directly connection with the
view of tunneling. Parikh and Wilczek \cite{pw00} calculated directly the
particle flux from the tunneling picture and made the tunneling physical
explanation holds firm basis. In their consideration the energy conservation
played a fundamental role and the outgoing particle itself created the barrier
\cite{pw04}. After this, there have been some works which have extended the
Parikh-Wilczek tunneling framework to different cases \cite{rzg05,jwc06} and
the question of information loss has been discussed in this framework
\cite{mv05,mv052}. Recently the general approach has been suggested
\cite{sk08} for the tunneling of matter from the horizon by using the first
law of thermodynamics or the conservation of energy. On the other side the
tunneling probability has also been calculated \cite{tp08} directly through
the change of the entropy that is proportional to area by the first law of
thermodynamics, which verifies the connection of black hole radiation with
thermodynamics \cite{swh76} further.

We have noticed that when the quantum gravity effect is considered the
tunneling formula can also be obtained by Parikh-Wilczek method and the
Hawking temperature relation \cite{amv05,ma060,ma06}. In this paper we will
proceed this kind of consideration by using the same method as in Ref.
\cite{tp08} but for the entropy which is modified by the logarithmic term
caused by quantum gravity effect as in Ref. \cite{amv05}. In the new method we
show clearly the necessary change of the surface gravity when considering the
quantum gravity effect and the crucial role which the Hawking temperature
relation plays. We note that the method could be extended to general situation
where the tunneling probability is obtained by calculating the change of
entropy, independent on the form of the entropy, from the first law of black
hole thermodynamics. The generalization verifies the connection of black hole
tunneling with thermodynamics further.

In this paper we take the unit convention $k=\hbar=c=G=1$.

\section{The first law of black hole thermodynamics and entropy}

The first law of black hole thermodynamics \cite{ht99} states:

If one throws a small amount of mass into a static non-charged and non-rotated
black hole, it will settle down to a new static black hole \cite{ep}. This
change can be described as $dM=\frac{\kappa}{8\pi}dA$, which is analogue to
the usual first law of thermodynamics $dM=TdS$. The case is the same for
radiating a small amount of mass from black hole \cite{swh76}.

According to Hawking, the temperature of black hole is taken as $T=\frac
{\kappa}{2\pi}$, so the entropy can be obtained as $S=\frac{1}{4}A$. It has
been shown \cite{tp08} that the tunneling formulas for static, spherically
symmetric black hole radiation are obtained by the first law of thermodynamics
and the area-entropy relation, even if the radiation temperature is different
from the Hawking temperature. From the first law of black hole thermodynamics,
we can see that if the black hole temperature is changed, the area-entropy
relation will also be changed. Note that in Ref. \cite{tp08} the author
calculated the tunneling probability by using the entropy being proportional
to horizon area and so the temperature was also proportional to the surface
gravity. But when considering the entropy which is modified by the logarithmic
term due to quantum gravity effect \cite{amv05}, it looks as if the black hole
temperature were not proportional to the black hole surface gravity. Then in
such situation, could the tunneling probability be obtained by calculating the
change of entropy with log-area term modification when considering the quantum
gravity effect in the same way as in Ref. \cite{tp08}? The answer is positive!
Before discussing this problem, we will first present the method proposed by Pilling.

\section{Thermodynamics and tunneling}

In this section we will review the method, presented in Ref. \cite{tp08},
which is used to obtain the tunneling probability directly from black hole
thermodynamics. Let us start by writing the metric for a general spherically
symmetric system in ADM form \cite{kw95},
\begin{equation}
ds^{2}=-N_{t}(t,r)^{2}dt^{2}+L(t,r)^{2}[dr+N_{r}(t,r)dt]^{2}+R(t,r)^{2}%
d\Omega^{2}.
\end{equation}
The metric is used for the situation where the geometry is spherically
symmetric and has a Killing vector which is timelike outside the horizon.
Specially one can consider the case of a massless particle and fix the gauge
appropriately ($L=1,R=r$) which is particularly useful to study across horizon
phenomena. So the metric becomes%

\begin{equation}
ds^{2}=-N_{t}(r)^{2}dt^{2}+[dr+N_{r}(r)dt]^{2}+r^{2}d\Omega^{2}, \label{me}%
\end{equation}
\ \ \ \ \ \ The metric is well behaved on the horizon and for a four
dimensional spherically Schwarzschild solution, $N_{t}=1,N_{r}=\sqrt{\frac
{2M}{r}}$ ($M$ is the mass of the black hole), for a four dimensional
Reissner-Nordstrom solution, $N_{t}=1,N_{r}=\sqrt{\frac{2M}{r}-\frac{Q^{2}%
}{r^{2}}}$ ($M$ is the mass and $Q$ is the charge of the black hole). And we
also note that for $N_{t}=\sqrt{\frac{f(r)}{g(r)}},N_{r}=f(r)\sqrt
{\frac{1-g(r)}{f(r)g(r)}}$, the metric (\ref{me}) becomes the same as that in
Ref. \cite{tp08}.

Now let us consider the Parikh-Wilczek tunneling \cite{pw00}. Supposed the
mass of the black hole is fixed and the mass is allowed to fluctuate, then the
shell of energy $E$ travels on the geodesics given by the line element
(\ref{me}). Taking into account self-gravitation effects, the outgoing radial
null geodesics near the horizon are given approximately by%

\begin{equation}
\overset{.}{r}=N_{t}(r)-N_{r}(r)\simeq(N_{t}^{^{\prime}}(R)-N_{r}^{^{\prime}%
}(R))(r-R)+O((r-R)^{2}), \label{rng}%
\end{equation}
where the horizon, $r=R$, is determined from the condition $N_{t}%
(R)-N_{r}(R)=0$ and the last formula is the expansion of the radial geodesics
in power of $r-R$.

According to the definition of a time-like Killing vector the surface gravity
of the black hole near the horizon is obtained as%

\begin{equation}
\kappa\simeq N_{t}^{^{\prime}}(R)-N_{r}^{^{\prime}}(R). \label{sg}%
\end{equation}
\ 

So the radiation temperature is%

\begin{equation}
T=\frac{\kappa}{2\pi}=\frac{N_{t}^{^{\prime}}(R)-N_{r}^{^{\prime}}(R)}{2\pi}.
\label{rt}%
\end{equation}

Now we consider the black hole thermodynamics in the region near the horizon.
The change of the Bekenstein-Hawking entropy, if the mass of black hole
changes from $M_{i}$ to $M_{f}$, is given as%

\begin{equation}
\Delta S=\int_{M_{i}}^{M_{f}}\frac{dS}{dM}dM=\int_{M_{i}}^{M_{f}}2\pi
R\frac{dR}{dM}dM. \label{bh}%
\end{equation}

Considering the small path near $R$, we can insert the mathematical identity
$\operatorname{Im}\int_{r_{i}}^{r_{f}}\frac{1}{r-R}dr=-\pi$ in the formula
(\ref{bh}). Thus we obtain%

\begin{equation}
\Delta S=-2\operatorname{Im}\int_{M_{i}}^{M_{f}}\int_{r_{i}}^{r_{f}}\frac
{R}{r-R}\frac{dR}{dM}dM. \label{bhe}%
\end{equation}

Using (\ref{rt}) and the expression of the temperature in thermodynamics
$\frac{1}{T}=\frac{\partial S}{\partial E}$, we attain%

\begin{equation}
R\frac{dR}{dM}=\frac{1}{N_{t}^{^{\prime}}(R)-N_{r}^{^{\prime}}(R)}. \label{ts}%
\end{equation}

Then the equations (\ref{rng}) and (\ref{ts}) give the final form of the
change of entropy (\ref{bhe}) as%

\begin{equation}
\Delta S=-2\operatorname{Im}\int_{M_{i}}^{M_{f}}\int_{r_{i}}^{r_{f}}\frac
{dR}{\overset{.}{r}}dM=-2\operatorname{Im}I,
\end{equation}
where $I$ is the action for an s-wave outgoing positive particle in WKB approximation.

So the tunneling probability is given as%

\begin{equation}
\Gamma\thicksim e^{\Delta S}=e^{-2\operatorname{Im}I}%
\end{equation}

Thus we obtain the tunneling probability from the change of entropy as a
direct consequence of the first law of black hole thermodynamics by using the
same method as that in Ref. \cite{tp08}. Let us emphasize that in the original
method the author uses the general radiation temperature different from the
Hawking temperature in order to discuss the factor of 2 problem. However the
new temperature is still proportional to the surface gravity like the Hawking
temperature and only the proportional relation is crucial for the discussed
problem in this paper. So we take the Hawking temperature as the black hole
temperature without loss of generality.

\section{The tunneling through the quantum horizon}

We note that for spherically symmetric black holes a generalized treatment
\cite{sk08} has been suggested, in which the tunneling probability is gotten
directly from the principle of conservation of energy by calculating the
imaginary part of the action in WKB approximation and the method is
independent on the form of black hole entropy. For the entropy which is
proportional to area \cite{pw00} or contains the logarithmic modification
caused by the presence of quantum gravity \cite{amv05}, we know that the
tunneling probability has been obtained by calculating the imaginary part of
the action in WKB approximation. Recently Pilling has suggested that the
tunneling probability is obtained directly from the first law of
thermodynamics by calculating the change of the entropy being proportional to
area, even if the radiation temperature is different from the Hawking
temperature \cite{tp08}. Then could the Pilling method be applied to the
situation where the entropy is modified by logarithmic term when considering
quantum gravity effect? In the following we will discuss the problem.

First we take into account the modification of entropy caused by the presence
of quantum gravity effect which gives a leading order correction with a
logarithmic dependence on the area besides reproducing the familiar
Bekenstein-Hawking linear relation \cite{dvf95,km00,dnp04}%

\begin{equation}
S_{QG}=\frac{A}{4L_{P}^{2}}+\alpha\ln\frac{A}{L_{p}^{2}}+O(\frac{L_{p}^{2}}%
{A}),
\end{equation}
where $A$ is the area of black hole horizon and $L_{p}$ is the Planck length.
The relation exists in string theory and loop quantum gravity. The difference
is that $\alpha$ is negative in the case of loop quantum gravity \cite{gm05},
but in String Theory the sigh of $\alpha$ depends on the number of field
species appearing in the low energy approximation \cite{sns98}. It is noted
that there is an interesting phenomenon that this log-area correction is
closely related to black hole remnant when the coefficient $\alpha$ is
negative \cite{lx07}.

Along Pilling's line we calculate the tunneling probability by using the
entropy modified by quantum gravity effect. For briefness, we write%

\begin{equation}
S_{QG}=\frac{1}{4}A+\alpha\ln A=\pi R^{2}+\alpha\ln(4\pi R^{2}).
\end{equation}
where the logarithmic correction can also be obtained by considering the
one-loop effects of quantum matter fields near a black hole \cite{dvf95,dnp04}%
. Whatever consideration we take, the spacetime will change. If we continue to
use the spacetime represented by (\ref{me}), the wrong result will be gotten.
We can see this point clearly from the following calculation.

When the mass of the black hole changes from $M_{i}$ to $M_{f}$, we have%

\begin{equation}
\Delta S=\int_{M_{i}}^{M_{f}}\frac{dS}{dM}dM=\int_{M_{i}}^{M_{f}}\left(  2\pi
R+\frac{2\alpha}{R}\right)  \frac{dR}{dM}dM=\Delta S_{1}+\Delta S_{2},
\label{qe}%
\end{equation}
where $\Delta S_{1}=\int_{M_{i}}^{M_{f}}2\pi R\frac{dR}{dM}dM$, $\Delta
S_{2}=\int_{M_{i}}^{M_{f}}\frac{2\alpha}{R}\frac{dR}{dM}dM$. It is noted that
if we continue to calculate according to the same method presented in the
section above, it will be found that $\Delta S_{1}=-2\operatorname{Im}%
\int_{M_{i}}^{M_{f}}\int_{r_{i}}^{r_{f}}\frac{dR}{\overset{.}{r}}dM$ by using
the surface gravity (\ref{sg}). It seems that $\Delta S_{2}$ is not related to
the action of the black hole and so is not related to the tunneling
probability. This is inconsistent with the result obtained in Ref.
\cite{amv05} where%

\begin{equation}
\Gamma(E)\thicksim e^{-2\operatorname{Im}I}=\left(  1-\frac{E}{M}\right)
^{2\alpha}e^{\left(  -8\pi ME(1-\frac{E}{2M})\right)  }.
\end{equation}
Consequently, we see that the imaginary part of the action is expressed as the
change of the whole entropy but not that of partial entropy. The calculation
above shows that when considering the entropy with logarithmic correction, the
spacetime will change and carry the quantum gravity effect. On the other side
we note that in Ref. \cite{tp08} the author uses the entropy being
proportional to area, so the radiation temperature is obviously proportional
to the surface gravity. But here we take the entropy $S_{QG}=\frac{1}%
{4}A+\alpha\ln A$, it looks as if formally the temperature were not
proportional to the surface gravity according to the first law of black hole
thermodynamics. A straight way to contain the quantum gravity effect is to use
the thermodynamic relation to get the surface gravity afresh. From
thermodynamics, the temperature can be given as%

\begin{equation}
\frac{1}{T}=\frac{dS}{dM}=\left(  2\pi R+\frac{2\alpha}{R}\right)  \frac
{dR}{dM}\equiv\frac{2\pi}{\kappa}.
\end{equation}

Thus we can get the surface gravity in the entropy with logarithmic
modification as%

\begin{equation}
\kappa\equiv2\pi/\frac{dS_{QM}}{dM}=\frac{2\pi}{\left(  2\pi R+\frac{2\alpha
}{R}\right)  \frac{dR}{dM}}, \label{qsg}%
\end{equation}
where the surface gravity is not only dependent on the mass of the black hole
but also dependent on the coefficient $\alpha$ which accords with the
consideration that the surface gravity carries the quantum gravity correction.

We note that if we want to obtain the relation $\Delta S=-2\operatorname{Im}I$
as that in the section above when considering the entropy with logarithmic
correction, we have to find the method to calculate the radial null geodesic
trajectory which is difficult to be calculated because we don't know the
property of such spacetime clearly. At the same time it has been pointed out
that the quantum entropy comes from counting states in a quantum theory,
whereas geodesics make sense in a classical spacetime. So the concept of
geodesics has to be managed carefully when the logarithmic correction of
entropy is explained as quantum gravity effect \cite{amv05,ma060,ma06}.
However, the attained result is consistent with the explanation of the
tunneling probability of quantum mechanics. Thus the feasibility of using the
concept of geodesics means that there maybe exist the physical reason to
explain the mathematical consistency. We note that the logarithmic correction
of the black hole entropy can be obtained from the purely quantum gravity
effect and can also be obtained from the one-loop effects of quantum matter
fields near a black hole \cite{dnp04}. The difference lies in the value of the
parameter $\alpha$, but the problem here is not concerned about it. So we can
calculate the geodesics by considering the one-loop effects of quantum matter
fields near a black hole. It is noted that in such consideration the
expression of spacetime presented in (\ref{me}) could still be used
\cite{dvf95,jwy85}, but some quantities, such as the mass, the temperature,
the surface gravity and so on, has to be changed. On the other hand we also
note that here the modification of surface gravity (\ref{qsg}) is consistent
with the result obtain by considering the one-loop correction as in Ref.
\cite{jwy85,ls88}. For example, for Schwarzschild spacetime, the classical
surface gravity is expressed as $\kappa_{0}=\frac{1}{4M}$ and the radius is
$R=2M$, so by Eq. (\ref{qsg}) the modified surface gravity can be gotten as
$\kappa\simeq$ $\kappa_{0}(1-\frac{\alpha}{4\pi M^{2}})$ which accords with
the modified surface gravity due to one loop back reaction effects
\cite{jwy85,ls88}.

In the following we will show that the tunneling probability can be recovered
by calculating the change of the entropy with logarithmic modification. By
(\ref{qsg}), we can write the change of entropy as%

\begin{equation}
\Delta S=\int_{M_{i}}^{M_{f}}\left(  2\pi R+\frac{2\alpha}{R}\right)
\frac{dR}{dM}dM=\int_{M_{i}}^{M_{f}}\frac{2\pi dM}{\kappa}. \label{ec}%
\end{equation}
\ \ \ \ \ \ \ \ \ \ \ \ 

Because the spacetime (\ref{me}) can still be used, so the radial null
geodesic trajectory is written as \cite{sp99,jw06},
\begin{equation}
\overset{.}{r}\simeq\kappa(r-R), \label{rngt}%
\end{equation}
where the formula can be gotten by using Eq. (\ref{rng}) and Eq. (\ref{sg})
and it must be stressed that here the surface gravity and the event horizon
have been changed and are different from that appeared in the section above.
Thus we replace the surface gravity in Eq. (\ref{ec}) by Eq. (\ref{rngt}),
insert the mathematical identity $\operatorname{Im}\int_{r_{i}}^{r_{f}}%
\frac{1}{r-R}dr=-\pi$ and have%

\begin{equation}
\Delta S=-2\operatorname{Im}\int_{M_{i}}^{M_{f}}\int_{r_{i}}^{r_{f}}\frac
{1}{\overset{.}{r}}drdM=-2\operatorname{Im}I.
\end{equation}

So the tunneling probability is gotten as%

\begin{equation}
\Gamma\thicksim e^{\Delta S}=e^{-2\operatorname{Im}\int_{r_{i}}^{r_{f}}%
p_{r}dr}. \label{tpc}%
\end{equation}
\ \ \ \ \ In this way we have finished the calculation of black hole tunneling
probability from the first law of thermodynamics when considering the quantum
gravity effect. In the derivation\ the introduction of the new surface gravity
is a crucial step because this maintains the general expression of the
relation between radiation temperature and surface gravity. We can see that
the Hawking temperature relation $T=\frac{\kappa}{2\pi}$ or the proportional
relation between the radiation temperature and surface gravity is key in the
calculation here as that in \cite{sk08,tp08,amv05}. In general, when we
discuss the connection of the black hole radiation as tunneling with
thermodynamics, only if we accept the Hawking temperature relation or the
proportional relation between the temperature and surface gravity, can we
obtain the tunneling probability directly from the first law of the black hole
thermodynamics, not dependent on the form of entropy of the black hole, which
is seen by writing the change of the entropy as%

\begin{equation}
\Delta S=\int_{M_{i}}^{M_{f}}\frac{dS}{dM}dM=\int_{M_{i}}^{M_{f}}\frac{1}%
{T}dM=\int_{M_{i}}^{M_{f}}\frac{2\pi dM}{\kappa}.
\end{equation}
\ Thus we can conclude that it is the relation between black hole temperature
and surface gravity that plays the crucial role that relates the black hole
thermodynamics with the tunneling picture of the black hole. At the same time
the concept of geodesics has to be managed carefully.

After the Hawking temperature was discovered, there have also been several
other methods \cite{hh76,sp99} to derive the same result as that obtained by
Hawking \cite{swh75}. Recently, however, it has been pointed out \cite{aas06}
that the tunneling approach produces a temperature that is double the original
Hawking temperature, which is used to question either the tunneling methods or
the value of Hawking temperature. This problem is discussed again in
\cite{pm07} where the authors consider the incoming solution besides the
outgoing solution and uses the ratio of the outgoing and incoming
probabilities to recover the Hawking temperature. The factor of 2 problem
about black hole temperature is also discussed generally in Ref. \cite{tp08}.

\section{\textbf{Conclusion}}

We have showed that the tunneling probability can be obtained from the first
law of thermodynamics by using the entropy with logarithmic modification which
contains the quantum gravity effect and the change of the surface gravity has
been presented clearly in the calculation. We have also showed the important
role that the relation between the radiation temperature and the surface
gravity plays. One can note that our derivation can be generalized only by
starting from the first law of thermodynamics $dM=TdS$ and the relation
$T=\frac{\kappa}{2\pi}$ instead of considering the form of the black hole
entropy. The generalization verifies the connection of black hole radiation
with thermodynamics further.

\section{Acknowledgement}

We are grateful to the anonymous referee for his/her critical comments and
helpful advice. This work is fund by National Natural Science Foundation of
China (Grant. No. 10504039 and 10747164).


\begin{thebibliography}{99}
\bibitem {ep}The law is also applicable to the general situation such as that
of charged and rotating black holes.
\end{thebibliography}
\end{document}